\documentclass{ifacconf}

\usepackage{graphicx}      
\usepackage{natbib}        
\usepackage[overload]{empheq}
\usepackage{amsmath,amsfonts}
\usepackage{mathtools}
\usepackage{bm}
\usepackage{xcolor}
\usepackage{tikz}
\usepackage{subcaption}
\usepackage{stmaryrd}
\usepackage{bm}
\definecolor{dgreen}{rgb}{0.0,0.5,0.0}
\usepackage{listings}
\usepackage{url}
\usepackage{cancel}

\newcommand{\bs}{\boldsymbol}

\newcommand*\rfrac[2]{{}^{#1}\!/_{#2}}
\newenvironment{review}{\color{black}}{}
\newcommand{\rev}[1]{\begin{review}#1\end{review}}
\begin{document}
\begin{frontmatter}

\title{PNLSS Toolbox 1.0\thanksref{footnoteinfo}} 

\thanks[footnoteinfo]{This work was supported by the Flemish fund for scientific research FWO under license number G0068.18N.}

\author[First]{Jan Decuyper}
\author[Second]{Koen Tiels} 
\author[First,Third]{Johan Schoukens}

\address[First]{Department of Engineering Technology, Vrije Universiteit Brussel, Pleinlaan 2, 1050 Brussels, Belgium (e-mail: jan.decuyper@vub.be, johan.schoukens@vub.be).}
\address[Second]{Department of Mechanical Engineering, Eindhoven University of Technology, Eindhoven, The Netherlands (e-mail: k.tiels@tue.nl)}
\address[Third]{Department of Electrical Engineering, Eindhoven University of Technology, Eindhoven, The Netherlands }


\begin{abstract}                
This is a demonstration of the PNLSS Toolbox 1.0. The toolbox is designed to identify polynomial nonlinear state-space models from data. Nonlinear state-space models can describe a wide range of nonlinear systems. An illustration is provided on experimental data of an electrical system mimicking the forced Duffing oscillator, and on numerical data of a nonlinear fluid dynamics problem.
\end{abstract}


\end{frontmatter}

\section{Introduction}
The PNLSS Toolbox 1.0\footnote{The toolbox is made available through \url{http://homepages.vub.ac.be/~jschouk/}} \citep{tiels2016} is designed to identify models describing nonlinear dynamical relationships. The generated models are of the \textbf{P}olynomial \textbf{N}on\textbf{L}inear \textbf{S}tate \textbf{S}pace type. The identification technique can be classified as fully black-box, meaning that the relationship between input(s) and output(s) is inferred solely from data.

Nonlinear state-space models can be seen as a universal model class, encompassing a wide range of popular model structures, i.a.\ Volterra series \citep{schetzen1980}, NARX models \citep{billings2013}, and block-oriented models \citep{schoukensM2017}.
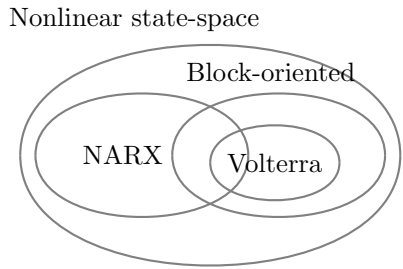
\begin{figure}[h]
\begin{center}
\begin{tikzpicture}
\draw[thick,gray] (0,0) to [out=90,in=90] (5,0);
\draw[thick,gray] (0,0) to [out=270,in=270] (5,0);
\node at (1.5,1.8) {Nonlinear state-space};

\draw[thick,gray] (0.2,0) to [out=90,in=90] (3,0);
\draw[thick,gray] (0.2,0) to [out=270,in=270] (3,0);
\node at (1.35,0) {NARX};

\draw[thick,gray] (2,0) to [out=90,in=90] (4.8,0);
\draw[thick,gray] (2,0) to [out=270,in=270] (4.8,0);
\node at (3.3,1.1) {Block-oriented};

\draw[thick,gray] (2.5,-0.1) to [out=90,in=90] (4.2,-0.1);
\draw[thick,gray] (2.5,-0.1) to [out=270,in=270] (4.2,-0.1);

\node [right] at (2.6,-0.1) {Volterra};
\end{tikzpicture}
\end{center}
\caption{Illustration of the subclasses of nonlinear state-space models.}
\label{f:diagram}
\end{figure}

The model structure is comprised of two equations: a state equation which describes the nonlinear evolution of the underlying state, and an output equation which relates the output(s) to the states. Both equations may be nonlinear. The toolbox uses a discrete-time model representation given by \citep{paduart2010}
\begin{subequations}
\label{e:PNLSS}
\begin{align}
&\bs{x}(k+1) = \bs{A} \bs{x}(k) +\bs{B} \bs{u}(k) + \bs{f}(\bs{x}(k),\bs{u}(k))\\
&\bs{y}(k) = \bs{C} \bs{x}(k) + \bs{D} \bs{u}(k) + \bs{g}(\bs{x}(k),\bs{u}(k)),
\end{align}
\end{subequations}
where $k=tf_s$ is the time index with $t$ the time and $f_s$ the sample frequency. The inputs to the model are denoted by $\bs{u} \in \mathbb{R}^m$, the outputs are given by $\bs{y} \in \mathbb{R}^p$, and $\bs{x} \in \mathbb{R}^n$ is the vector of intermediate state variables, resulting in a model order of $n$. The functions $\bs{f}$ and $\bs{g}$ are multivariate vector functions, expressed in a polynomial basis,
\begin{subequations}
\begin{align}
&\bs{f}(\bs{x},\bs{u}) = \bs{E} \bs{\zeta}(\bs{x},\bs{u})\\
&\bs{g}(\bs{x},\bs{u}) = \bs{F} \bs{\eta}(\bs{x},\bs{u}),
\end{align}
\end{subequations}
where $\bs{\zeta} \in \mathbb{R}^{n_{\zeta}}$ and $\bs{\eta} \in \mathbb{R}^{n_{\eta}}$ are vectors of monomial basis functions. The basis expansion contains all possible cross-products between state variables and input variables raised to a user defined total power $d$, e.g.\ for $m=1$, $n=2$, and $d=2$ one would have
\begin{equation}
\bs{\zeta}(\bs{x},\bs{u}) = \left[ x_1^2\quad x_1x_2 \quad x_1u \quad x_2^2 \quad x_2u \quad u^2 \right]^{\text{T}}
\end{equation}
The matrices of coefficients then have the following dimensions: $\bs{A} \in \mathbb{R}^{n \times n}$, $\bs{B} \in \mathbb{R}^{n \times m}$, $\bs{C} \in \mathbb{R}^{p \times n}$, $\bs{D} \in \mathbb{R}^{p \times m}$,  $\bs{E} \in \mathbb{R}^{n \times n_{\zeta}}$, and $\bs{F} \in \mathbb{R}^{n \times n_{\eta}}$. The elements in these matrices are the parameters of the model.

Identifying a PNLSS model involves 3 steps:
\begin{enumerate}
\item Estimate a non-parametric linear approximation of the system.
\item Use linear frequency domain subspace identification to parameterise the linear estimate. 
\item Minimise the output error of the nonlinear model using nonlinear optimisation.
\end{enumerate}

The toolbox provides all the necessary utilities to complete the entire identification procedure starting from generating rich excitation signals up to model validation. The sections below provide a brief overview of the functionalities.  

\section{Signal generation}
\label{s:signal}

Rich excitation signals are imperative for data-driven modelling. The toolbox therefore contains a utility which enables the generation of random-phase multisine signals. A random-phase mutlisine signal $u(t)$ consists of a sum of harmonically related sine waves,
\begin{equation}
u(t) = \sum_{k=1}^{\lfloor N/2 \rfloor} A_k \sin \left( 2 \pi k \frac{f_s}{N} t + \phi_k \right)
\end{equation}
where $N$ is the number of samples, $f_s$ is the sampling frequency, and $A_k$ and $\phi_k$ are the amplitude and the phase of the sine wave with frequency $k\frac{f_s}{N}$. The phases are drawn from a uniform distribution in the interval from zero to $2\pi$. Multisine signals have proven highly effective for identification purposes given the reduced measurement times following from the persistent excitation over the required frequency band, and straightforward handling in the frequency domain given the periodicity. A particular choice of amplitudes, moreover, allows for a detection of nonlinear behaviour \citep{pintelon2012}.

The following code can be used to generate $R$ realisations of an odd random-phase multisine (exciting only the odd frequency lines) with a flat amplitude spectrum up to 90\% of the Nyquist frequency.
\begin{figure}[h]
\begin{center}
\includegraphics[width=0.48\textwidth]{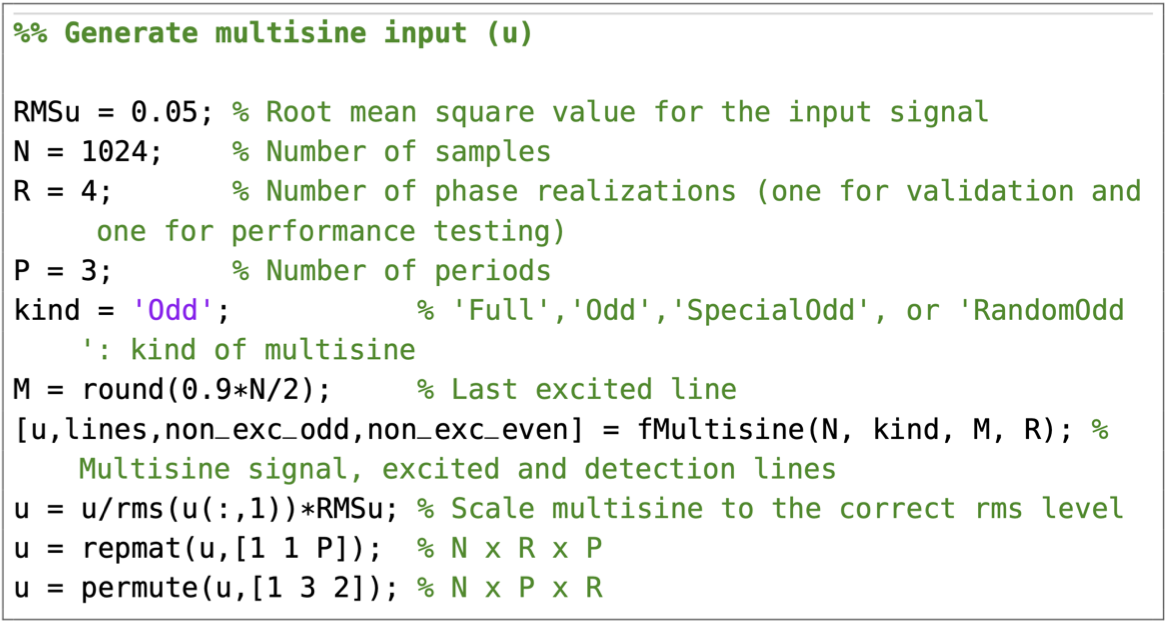}
\end{center}
\end{figure}

\section{Non-parametric frequency response estimation}
\label{s:BLA}

The first step in the identification procedure is the estimation of a non-parametric linear model (a frequency response function or FRF). The toolbox uses an implementation of the \emph{robust} method, which leads to the so-called \emph{Best Linear Approximation} (BLA) \citep{pintelon2012}.

The function \textbf{fCovarY} can be used to compute the FRF estimate \textbf{G}. Besides \textbf{G}, also the total (noise + nonlinear) distortion level, \textbf{covGML}, and the noise distortion level, \textbf{covGn}, on the FRF are computed. An estimate of the level of nonlinear distortion can be used as metric to assess the need for a nonlinear model. Using \textbf{fCovarY} the noise covariance on the output spectrum can be computed.
\begin{figure}[h]
\begin{center}
\includegraphics[width=0.48\textwidth]{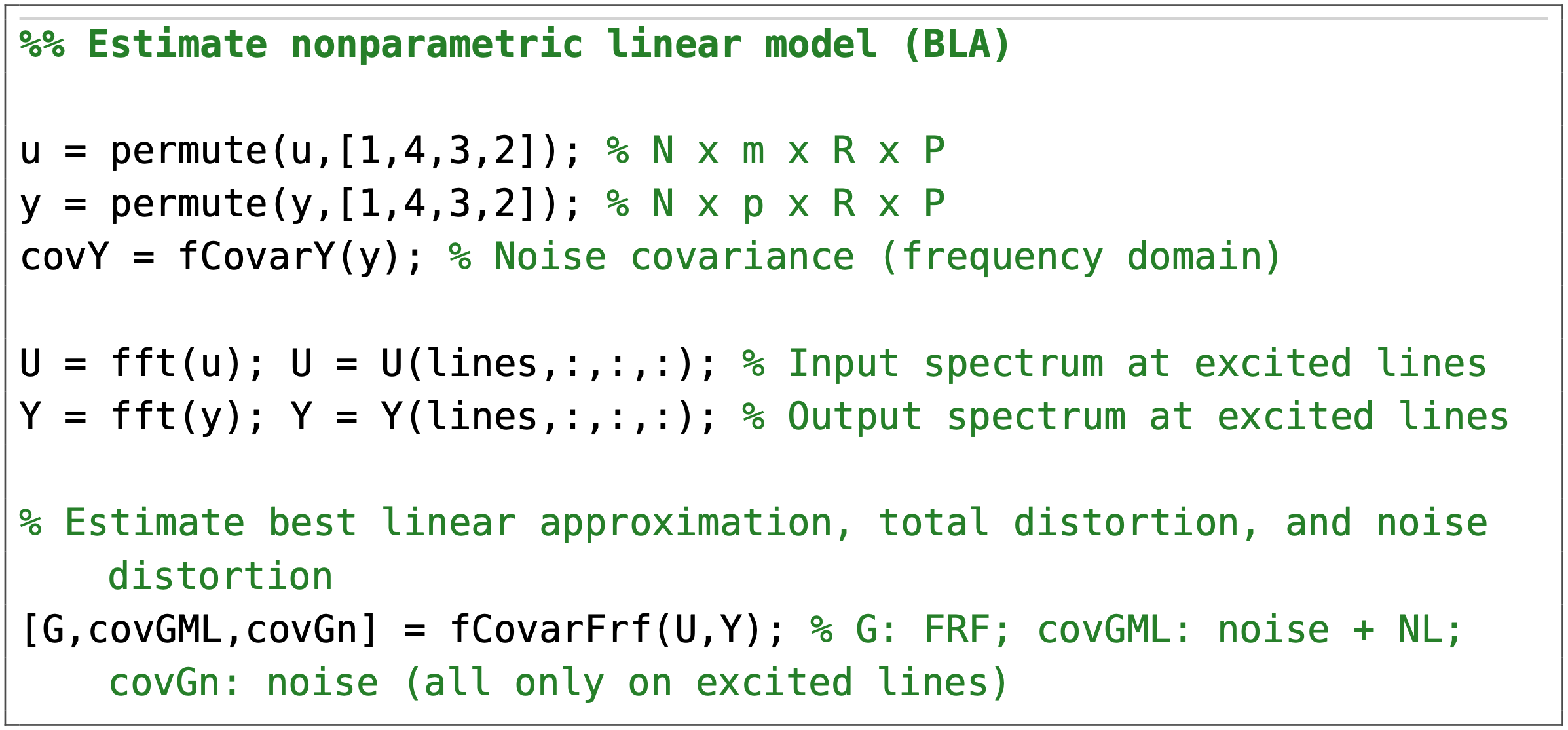}
\end{center}
\end{figure}

\section{Parametric linear subspace identification}
\label{s:subspace}

The goal is to initialise the nonlinear PNLSS model of Eq.~\eqref{e:PNLSS} using a parametric linear model. The non-parametric FRF estimate is therefore parametrised into a state-space model using frequency domain subspace identification \citep{mckelvey1996,pintelon2002}. The subspace method provides good initial estimates of the linear model parameters, i.e.\ $\bs{A},\bs{B},\bs{C},$ and $\bs{D}$. To further improve the linear estimate, a Levenberg-Marquardt optimisation is used. The total distortion level of Section \ref{s:BLA} is used as a frequency weighting in the optimisation. 

The code snippet below illustrates the use of the \textbf{fLoopSubSpace} function in computing parametric state-space models of the orders stored in $na$. In this case 100 iterations of the Levenberg-Marquardt optimisation were used. A particular model of order $n$ may then be selected on the basis of the accuracy of the fit, or based on cross-validation with a validation data set. 
\begin{figure}[h]
\begin{center}
\includegraphics[width=0.48\textwidth]{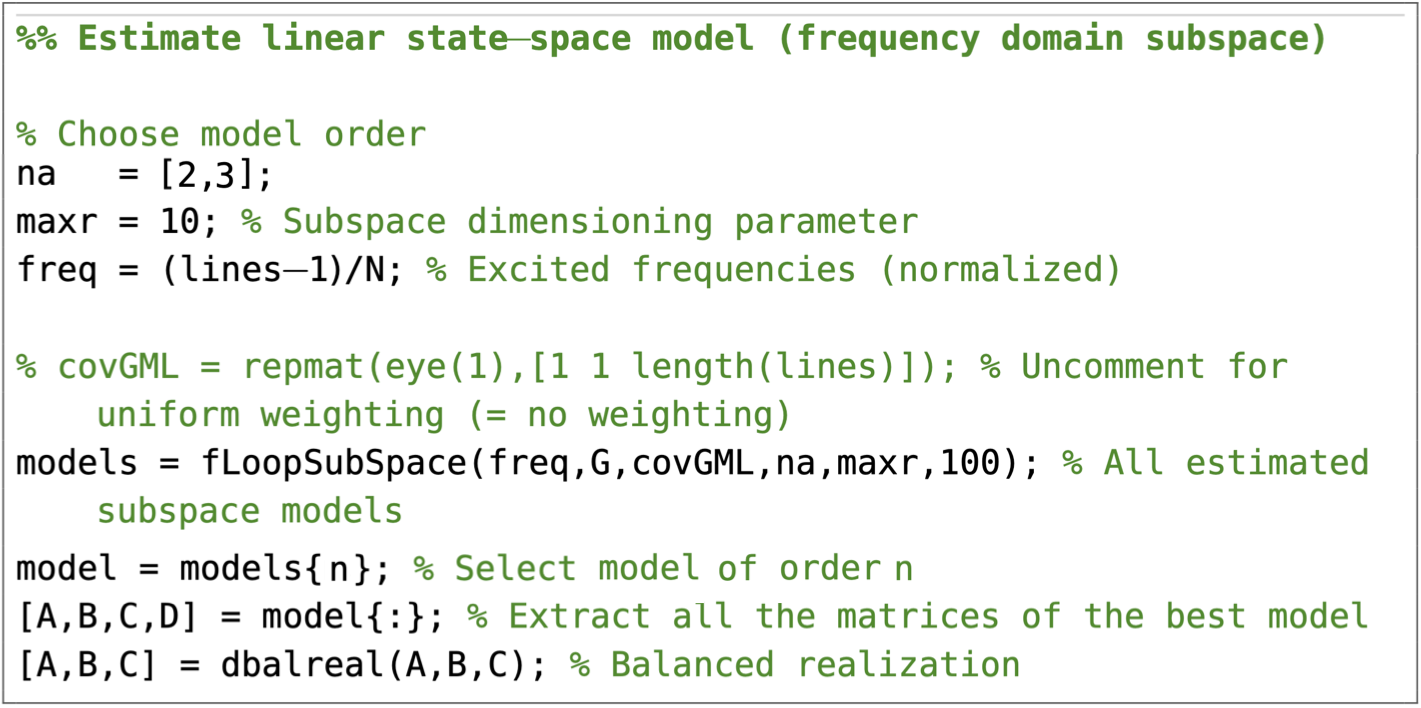}
\end{center}
\end{figure}

\section{Nonlinear model optimisation}
\label{s:LM}

The full nonlinear model is trained using nonlinear optimisation on all the matrices of coefficients. To obtain a compact training data set, it is common practice to first average over the periods of the signal. The training data is then formed out of a concatenation of the different experiments (phase realisations).
\begin{figure}[h]
\begin{center}
\includegraphics[width=0.48\textwidth]{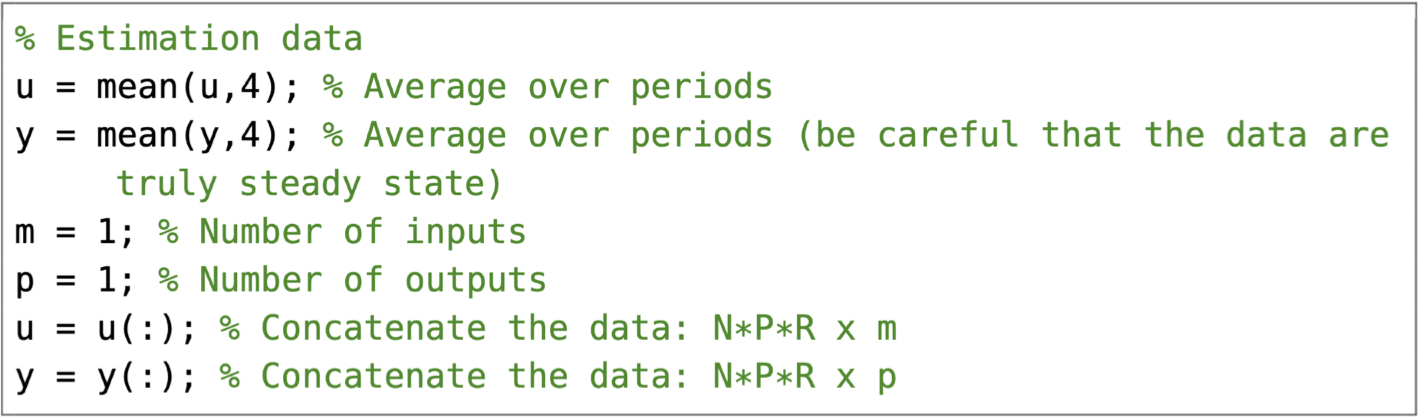}
\end{center}
\end{figure}

Concatenating multiple experiments will give rise to transient behaviour at the interconnections. To avoid that transients influence the parameter tuning, an additional period of each experiment can be \rev{artificially} prepended to the corresponding segment. The prepended period, \rev{which was not part of the original experiment,} then accommodates the transients and will be discarded when computing the cost. The number of samples to prepend, and the indices where to introduce them are stored in the vector $T_1$. In case of non-periodic data, a number of transient bearing samples of the original signal can be discarded. This number of samples is specified in the parameter $T_2$.

\begin{figure}[h]
\begin{center}
\includegraphics[width=0.48\textwidth]{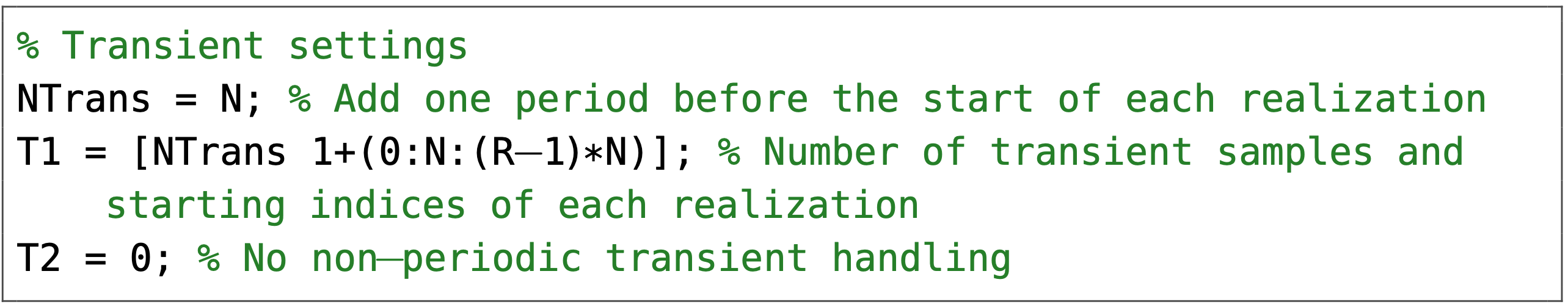}
\end{center}
\end{figure}

Using the linear initialisation of Section \ref{s:subspace} and by initialising the $\bs{E}$ and $\bs{F}$ matrices of the nonlinear part with zeros, a PNLSS structure of the form of Eq.~\eqref{e:PNLSS} is constructed. 
\begin{figure}[h]
\begin{center}
\includegraphics[width=0.48\textwidth]{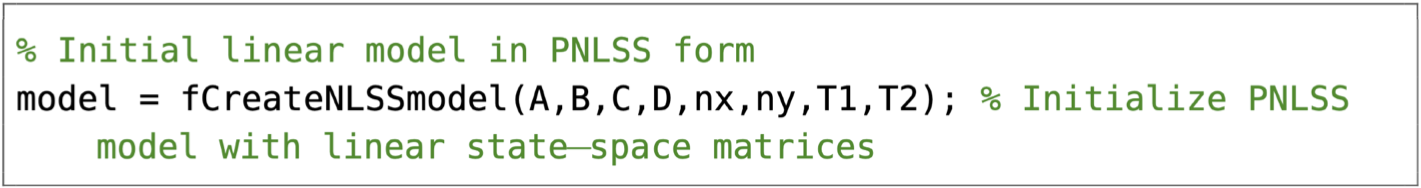}
\end{center}
\end{figure}

The arguments $n_x$ and $n_y$ are vectors storing the desired total powers of the monomials in the state and the output equation, respectively. For example, $n_x = [2,3]$ indicates quadratic and cubic polynomials in the state equation. Using the function \textbf{fSelectActive} one can select which basis functions to allow in the multivariate polynomials $\bs{f}$ and $\bs{g}$. A particular choice could be to only activate the monomials which are made up purely of state variables, e.g.\
\begin{equation}
\bs{\zeta}(\bs{x},\bs{u}) = \left[ x_1^2\quad x_1x_2 \quad \cancel{x_1u} \quad x_2^2 \quad \cancel{x_2u} \quad \cancel{u^2} \right]^{\text{T}}.
\end{equation}
Only the coefficients corresponding to active basis functions will be tuned. In the following example, the `statesonly' option was selected for the state equation while no nonlinear function was added to the output equation.

\begin{figure}[h]
\begin{center}
\includegraphics[width=0.48\textwidth]{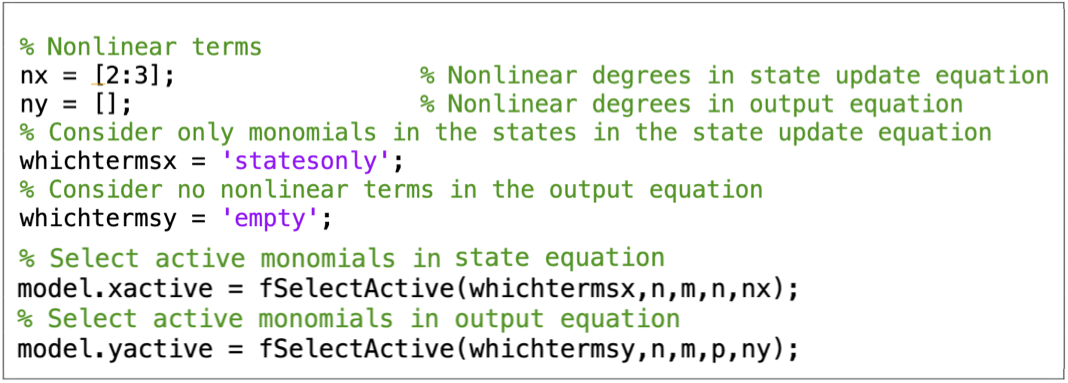}
\end{center}
\end{figure}

The optimisation is performed using a Levenberg-Marquardt algorithm. As stopping criterion a maximum number of iterations is used. Also an initial value of $\lambda$, a parameter balancing the Levenberg-Marquardt algorithm between gradient descent and Gauss-Newton needs to be set. This parameter will be updated after each iteration, i.e.\ when the step was successful $\lambda_{\text{new}} = 0.5\lambda$, leaning more towards the fast convergence of Gauss-Newton, when unsuccessful $\lambda_{\text{new}} = \lambda \sqrt{10}$, exploiting the large region of convergence of gradient descent.

If desired, a frequency-domain weighting may be applied. In this example the inverse of the noise covariance is used as weighting. It should however be stressed that in the likely event that structural model errors dominate (e.g.\ when the underlying system falls outside the model class), it is best practice to use a uniform weighting, presuming that the model errors are uniformly spread over the frequencies \citep{stoev2016}.
\begin{figure}[h]
\begin{center}
\includegraphics[width=0.48\textwidth]{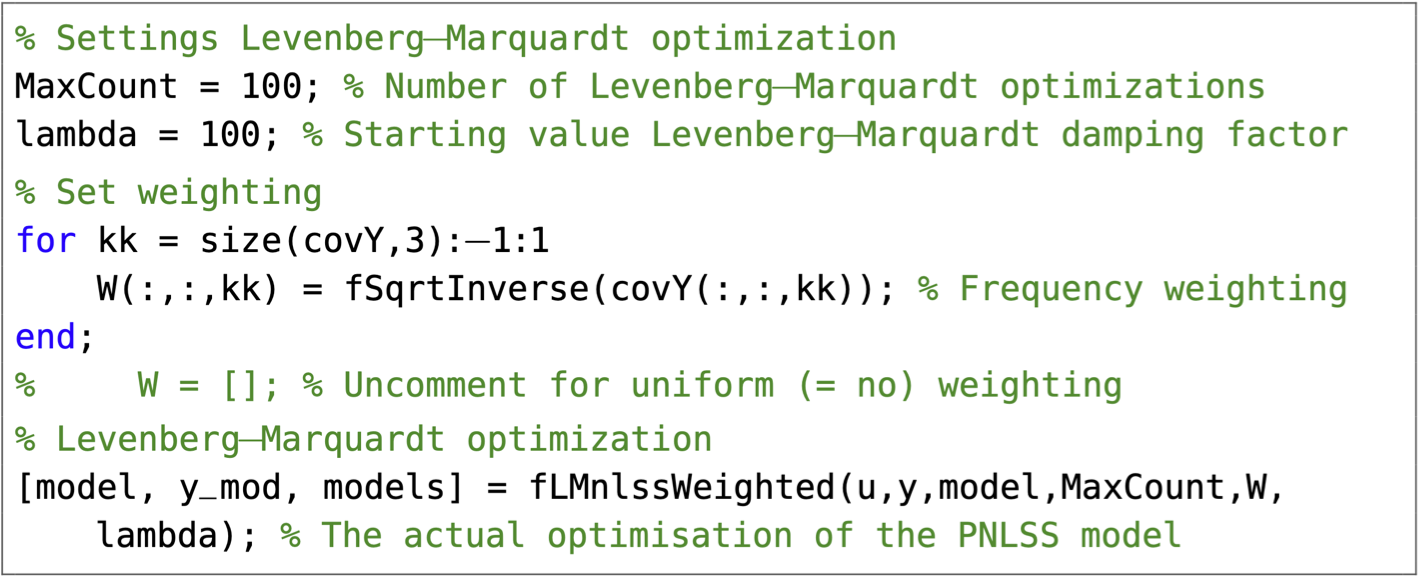}
\end{center}
\end{figure}

As an additional output, all the models which were obtained along the optimisation path are stored under `models'. Out of these models, the model which performs best on a cross-validation test can be selected.

To simulate the output to a new data set the function \textbf{fFilter\_NLSS} may be used.
\begin{figure}[h]
\begin{center}
\includegraphics[width=0.48\textwidth]{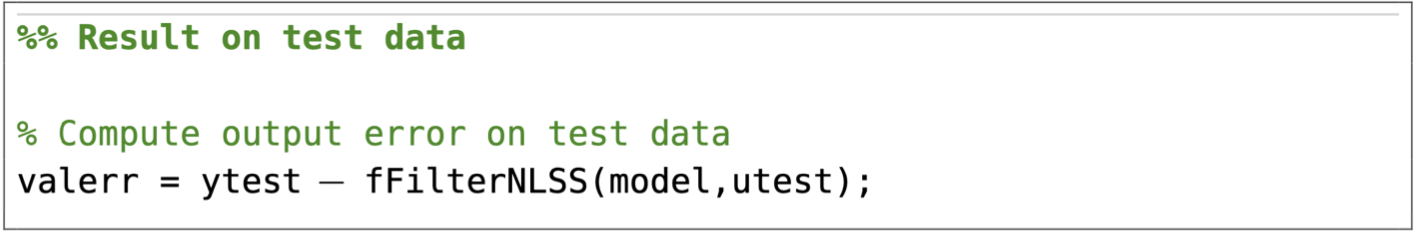}
\end{center}
\end{figure}

\section{Case studies}
\label{s:cases}

The PNLSS toolbox will be demonstrated on two data sets: one experimental data set describing the behaviour of a nonlinear circuitry, and one obtained from computational fluid dynamics simulations (CFD) describing a nonlinear fluid-structure interaction.

\subsection{The forced Duffing oscillator}
The data is obtained from an electrical implementation of a mechanically resonating system involving a moving mass $m$, a viscous damping $c$ and a nonlinear spring $k(y(t))$. The analogue electrical circuitry generates data close to but not exactly equal to the idealised representation given by the nonlinear ordinary differential equation (ODE)
\begin{equation}
\label{e:diff_SB}
m \ddot{y}(t) + c \dot{y}(t) + k(y(t))y(t) = u(t),
\end{equation}
where the presumed displacement, $y(t)$, is considered the output and the presumed force, $u(t)$, is considered the input. Overdots denote the derivative with respect to time. The static position-dependent stiffness is given by
\begin{equation}
\label{e:SB_kNL}
k(y(t))=\alpha+\beta y^2(t),
\end{equation}
which can be interpreted as a cubic hardening spring.

\textbf{The training data} consists of 9 realisations of a random-phase odd multisine.
The period of the multisine is $\rfrac{1}{f_0}$ with $f_0= \rfrac{f_s}{8192}$ Hz and $f_s\approx 610$ Hz. The number of excited harmonics is $L=1342$ resulting in an $f_{\text{max}}\approx 200$ Hz. Each multisine realisation is given a unique set of phases that are independent and uniformly distributed in $[0,2\pi[$. The signal to noise ratio at the output is estimated at approximately 40 dB. 

As \textbf{validation data} a filtered Gaussian noise sequence of the same bandwidth and with a linearly increasing amplitude is used. The data are part of three benchmark data sets for nonlinear system identification described in \cite{wigren2013} and are used in a number of works among which \cite{noel2018,ljung2004}.

A PNLSS model with the following characteristics is estimated: $m =1$, $n=2$, $p=1$, $n_x = [2,3]$, $n_y = [2,3]$. All monomials, constructed purely out of state variables, are allowed in the state equation while no nonlinear function is added in the output equation (i.e. $\bs{F}$ contains zeros). The PNLSS model is very accurate. It yiels a relative root-mean-squared error of only 0.5\% on the training data.

\begin{figure}[h]
\begin{center}
\includegraphics[width=0.45\textwidth]{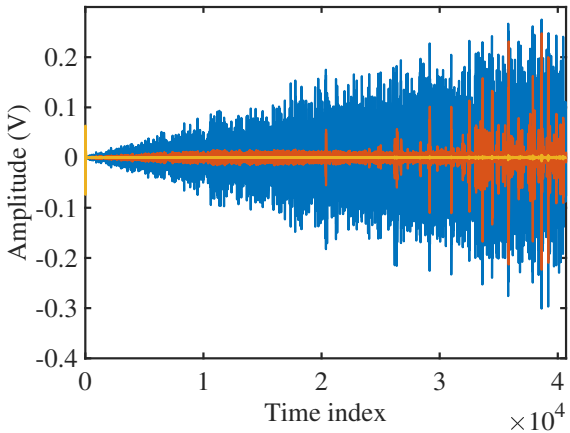}
\caption{Validation results of the forced Duffing model. Blue is the true data, red is the error of the linear model and yellow is the PNLSS error.}
\label{f:SB_test}
\end{center}
\end{figure}

Fig.~\ref{f:SB_test} shows the validation result of the model. The validation data extents beyond the amplitude level which was observed during training. It is clear that the linear model fails to extrapolate in the high-amplitude region while the PNLSS model remains accurate throughout the sequence. An overal relative root-mean-squared error level of 2\% is attained for the PNLSS model, whereas the linear model attains a relative rms error of 26\%.

\subsection{Modelling nonlinear fluid dynamics}
\label{s:VIV}

\begin{figure}[h]
\begin{center}
\includegraphics[width=0.45\textwidth]{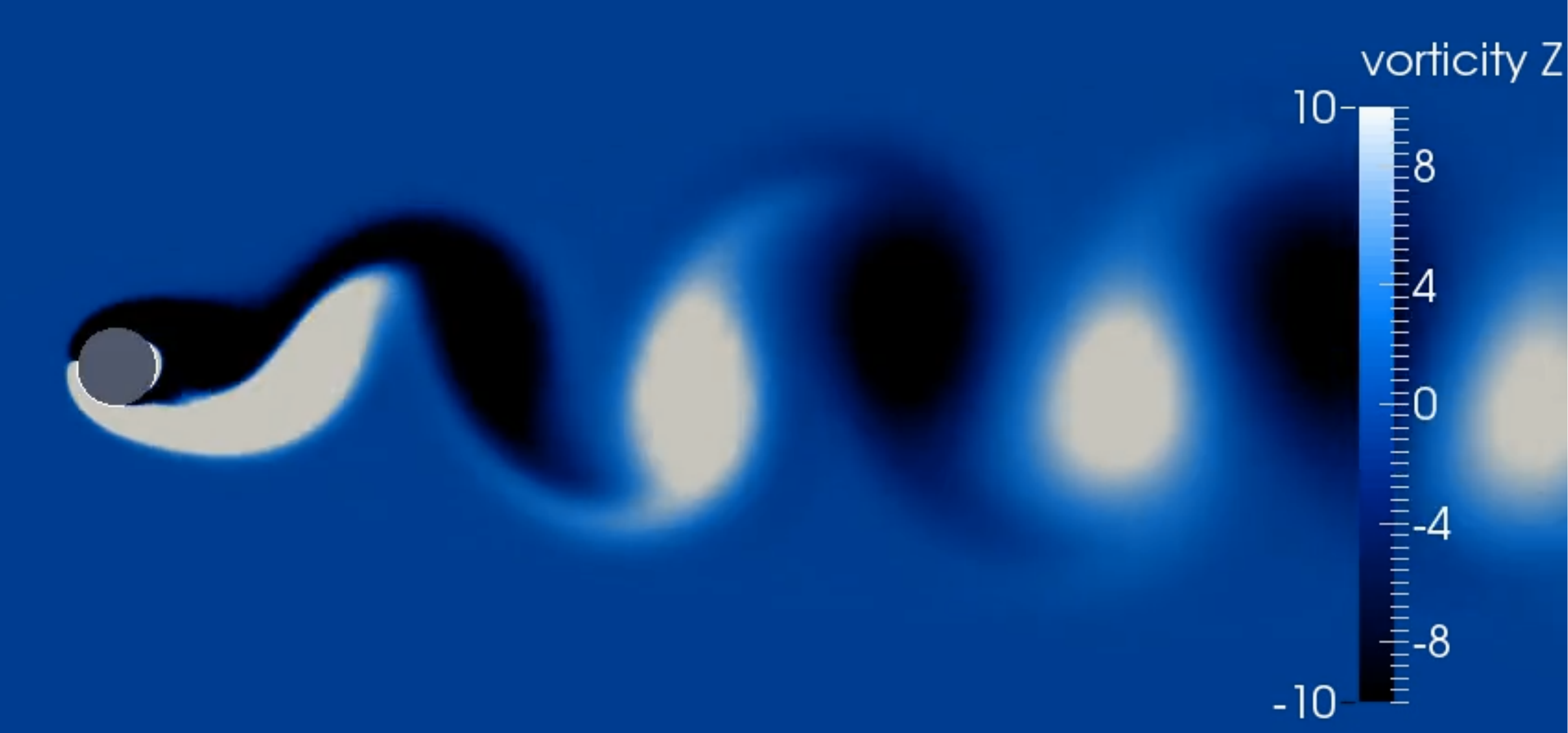}
\caption{CFD simulation of the vortex shedding in the wake of a 2D circular cylinder.}
\label{f:VIV}
\end{center}
\end{figure}

In \cite{decuyper2020} the forces that arise on a submerged cylinder in a uniform flow were modelled. The fluctuating forces originate back from an unsteady wake, characterised by alternating vortices. The phenomenon is of particular interest in civil engineering given the frequent encounter of slender, cylindrical shapes in the built environment. It is known that the displacement to force relationship is highly nonlinear, exhibiting limit cycles and synchronisation behaviour. An illustration of the setup is depicted in Fig.~\ref{f:VIV}.

\rev{It is common practice to study vortex-induced forces from forced oscillation experiments \citep{hover1997}. In that case the input corresponds to the imposed displacement of the cylinder while the output is the force coefficient, measured in the perpendicular direction to the oncoming flow.} 

\begin{figure}
\begin{center}
\includegraphics[width=0.47\textwidth]{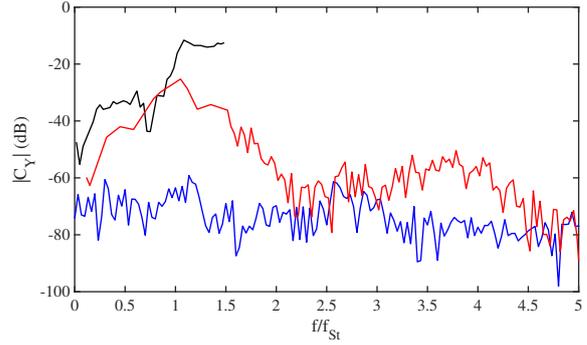}
\caption{Non-parametric nonlinear distortion analysis. Black depicts the output at the excited frequency lines. Blue indicates even nonlinear behaviour, observed at the unexcited even frequency lines. Red indicates odd nonlinear behaviour, observed at the unexcited odd frequency lines.}
\label{f:FAST}
\end{center}
\end{figure}

\rev{To study the nature of the nonlinearity, the system is first subjected to a non-parametric nonlinear distortion analysis. Using the signal generator functionality of the toolbox, random-phase random-odd multisine singals are generated. This particular flavour of multisines enables to both quantify the level of nonlinear distortions as well as label the behaviour as odd- or even-nonlinear \citep{pintelon2012}. A colour-coded output spectrum then exposes the system properties. Fig.~\ref{f:FAST} depicts the results for the unsteady fluid system. The system is found to behave dominantly odd-nonlinear. This information will be exploited when constructing the nonlinear model. The level of nonlinearity is significant, reaching as high as the signal itself at certain frequency lines.}

\rev{To train the model parameters,} data of a concatenation of swept sine signals are used. \rev{The sweeps cover a frequency band from 0 Hz up to $1.5~ \times$ the Strouhal frequency (i.e.\ the natural vortex shedding frequency, here 3 Hz). This ensures that both non-synchronised and synchronised vortex shedding regimes are captured.}   The data were generated using 2D computational fluid dynamic simulations. 

A PNLSS model with the following characteristics is estimated: $m =1$, $n=5$, $p=1$, $n_x = [0,3,5,7]$, $n_y = [0,3,5,7]$. \rev{Notice that the nonlinear degrees are tailored to the system based on knowledge of the nonlinear distortion analysis, which indicated dominantly odd-nonlinear behaviour (Fig.~\ref{f:FAST})}. All monomials were used both in the state and the output equation. The PNLSS model is able to accurately capture the training data resulting in a relative root-mean-squared error of only 4\%.

\begin{figure}[h]
\begin{center}
\includegraphics[width=0.45\textwidth]{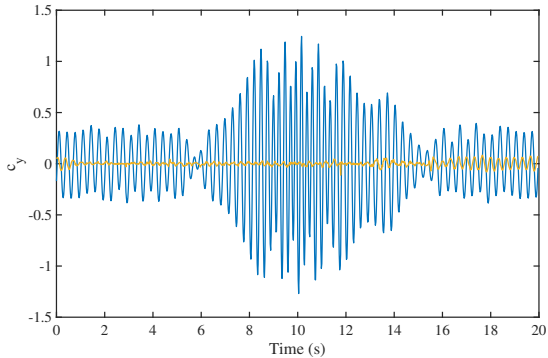}
\caption{Validation result of the fluid dynamics model. Blue is the true data and yellow is the PNLSS error.}
\label{f:VIV_val}
\end{center}
\end{figure}

As validation, the model was subjected to a new swept sine experiment. The result is shown in Fig.~\ref{f:VIV_val}. A low relative root-mean-squared error of only 6\% is retrieved.

\rev{The present case study illustrates that PNLSS models can be used to generate accurate reduced order models of highly nonlinear behaviour, typically governed by nonlinear partial differential equations. The obtained model is a local approximation of the system, tuned on a particular region of interest of the phase space.}

\section{Outlook}

Future versions of the toolbox will allow the use of external modules, e.g.\ the Matlab Optimisation Toolbox. Furthermore, the choice of basis function will be extended to non-polynomial bases. Additionally, it will be possible to construct nonlinear functions out of model outputs, such that $\bs{f}(\bs{x},\bs{u},\bs{y})$, and $\bs{g}(\bs{x},\bs{u},\bs{y})$. 

\rev{A well known hazard, frequently encountered during the parameter estimation of dynamical models, is the issue of exploding gradients (also referred to as model instabilities). The practical concerns related to dealing with non-contractive models may be addressed using multiple shooting techniques \citep{decuyper2020IFAC}, which will be included in an updated version of the toolbox.}

Recent advances in `decoupling techniques' have shown great potential for model reduction of nonlinear state-space models \citep{decuyper2021,decuyperFCPD2021}. \rev{It was found that highly structured representations of the nonlinearity could be retrieved starting from the classical monomial basis expansion. This alludes to the fact that nonlinear dynamical systems are in many cases driven by a low number of internal nonlinearities. The retrieval of underlying structure can lead to insight, which in turn contributes to bridging the gap between fully black-box modelling and physical modelling.} It is intended to include decoupling tools in the next version of the toolbox.  
\rev{
\section{Conclusion}
The PNLSS toolbox provides all required functionalities to be able to identify nonlinear state space models of nonlinear dynamical systems from data. The generic set of equations, featuring flexible polynomial nonlinearities, allow for a large variety of nonlinear systems to be modelled. An illustration based on two case studies, i.e.\ experimental data obtained from the forced Duffing oscillator and numerical data of an unsteady fluid dynamics phenomenon, was presented.
}
\bibliography{entirebib} 
\end{document}